\documentclass{aa}
\usepackage{graphicx}
\usepackage{txfonts}

\newcommand{\lya}{\mbox{Ly$\alpha$}}

\newcommand{\OIII}{\mbox{[O\thinspace{\sc iii}]}}

\begin{document}
   \title{Reanalysis of the spectrum of the $z=10$ galaxy}


   \author{S. J. Weatherley
             \and
          S. J. Warren         
          \and
          T. S. R. Babbedge
          }

   \offprints{S. J. Weatherley, \email{stephen.weatherley@imperial.ac.uk}}

   \institute{Astrophysics Group, Blackett Laboratory, Imperial College London, London SW7 2BW, UK
                    }


   \abstract{
   In a recent paper Pell\'{o} et al. reported observations of a faint
galaxy, gravitationally lensed by the galaxy cluster Abell 1835. Deep
J--band spectroscopy revealed a weak emission line near 1.34 microns,
detected in two spectra with different central wavelengths. The line
was interpreted as \lya\ at redshift $z=10.0$. This interpretation is supported
by the broad--band photometric spectral energy distribution, and by
the location of the galaxy close to the lens critical line for this
redshift. We have reanalysed the two spectra, just released from the
data archive. Our analysis includes allowance for wavelength shifts
due to transverse drift of the object in the slit. We do not detect a 
significant emission line at the
reported location, or nearby, at either grating setting, nor in the combined
spectrum. We provide a possible explanation for the reported detection as
due to spurious positive flux introduced in the sky--subtraction stage as a result of
variable hot pixels. We provide our final reduced 2D frame, and corresponding
error array.
   }

   \maketitle
%

\section{Introduction}

  The new generation of telescopes and instruments have allowed
astronomers to explore beyond redshift $z=6$. Analysis of the spectra
of the highest--redshift quasars discovered (Fan et al., 2003),
indicate that we may be on the threshold of the epoch at which the
intergalactic medium was reionised. The epoch of reionisation is
predicted to occur over a relatively short redshift interval (Gnedin,
2000). As such it is seen as a fundamental event in cosmic history,
and the study of this epoch is one of the great goals of observational
cosmologists. Analysis of the WMAP one--year polarisation cross--power
spectra (Kogut et al., 2003) indicates a higher redshift of
reionisation than the quasar data, $11<z_r<30$. This may point to a
more complex history of reionisation than previously predicted. These
results motivate searches for galaxies and quasars beyond $z=7$, to
measure the conditions in the intergalactic medium at these
times. Quasars are particularly useful, because bright, but are
expected to be extremely rare, requiring very ambitious surveys to
find any (Warren and Hewett, 2002). The galaxy population at such high
redshifts will comprise objects of low mass, and therefore low
luminosity, which, coupled with the large distances, and the bright
near--infrared sky, again make detection a considerable
challenge. More fundamentally, the expected high optical depth of the
neutral fraction of the intervening intergalactic medium could
obliterate the \lya\ line, the principal means of redshift confirmation
(Miralda--Escud\'{e}, 1998). In this context, at a time when
there are no convincing spectroscopically--confirmed detections of
galaxies beyond $z=7$, the publication of the detection of \lya\ from
a galaxy at $z=10.0$ by Pell\'{o} et al. (2004; hereafter P04), is of
great interest.

The galaxy detected by Pell\'{o} et al., labelled $\#1916$, lies behind
the galaxy cluster Abell 1835. The galaxy is detected in the
near--infrared H and K bands, but shows a sharp drop in flux in moving
into the J band, and is not detected in very deep observations in the
optical R and I bands. This spectral energy distribution is consistent
with models for young galaxies in the redshift interval $9<z<11$,
where the spectral break is a consequence of complete absorption
blueward of redshifted \lya\, by neutral gas in the intergalactic
medium. This interpretation is supported by the location of the galaxy
close to the lens critical line for these redshifts. For these reasons
Pell\'{o} et al. obtained spectra in the J band, selecting a range of
central wavelengths to target \lya\ over the indicated redshift
interval. They reported the detection of a weak emission line at
wavelength $1.33745\mu$m. The line appeared in two spectra, in the
region of overlap of two wavelength intervals. The flux is
$(4.1\pm0.5)\times10^{-18}$\,erg\,s$^{-1}$\,cm$^{-2}$ i.e. $8\sigma$
(we understand that the `$4-5\sigma$' quoted in the paper refers to
the peak pixel (Pell\'{o}, private communication)). They
identified the line as \lya\ at $z=10.0$.
In the published spectrum (fig. 5 in P04), comparison of the line flux
with the $1\sigma$ error spectrum appears to indicate a higher significance
than the quoted $8\sigma$. Curious to
understand this inconsistency, we downloaded the raw data from the ESO
archive, when they became publicly available on 2 July 2004. In this
paper we present a reanalysis of the two spectra. We do not detect the
emission line in either spectrum. In \S2 we detail the data reduction
steps followed, and in \S3 we present the results of analysing the
reduced frames, and briefly discuss possible explanations for the
discrepancy between our results and those published in P04.

\section{Data Reduction}
Details of the observations, obtained with the ISAAC instrument on the
ESO--VLT, are provided in P04. Briefly, the data comprise 12 frames of
900s integration at the first grating setting, central wavelength
$\lambda=1.315\mu$m (hereafter $\lambda1$), and 21 frames of 900s
integration at the second grating setting, central wavelength
$\lambda=1.365\mu$m (hereafter $\lambda2$). The particular region of interest for this 
paper is the region of overlap: the wavelength range 1.3357 - 1.3409 microns.
The $\lambda2$
observations were taken over two nights; 9 frames on the first\footnote{This sequence was terminated due to worsening conditions.},
and 12 on the second. The data were reduced in three sets
corresponding to these groupings, then the two groups at $\lambda2$
were combined. The slit width was $1\arcsec$, and the seeing
$0.4\arcsec$ to $0.6\arcsec$. The wavelength range of each spectrum is
$0.059\mu$m, with 1024 pixels, and a spatial pixel scale of
$0.148\arcsec$. The nominal resolving power for an object filling the
slit is 3100, corresponding to approximately 7\thinspace pixels. However, for a
point source, because of the good seeing, the expected line width for
a spectrally unresolved line becomes 3 or 4 pixels.

Because of the moderately high dispersion, in regions between bright
OH sky lines the noise in an individual frame is dominated by
detector noise, and not photon noise. The detected emission line lies
in such a region; therefore, our data reduction procedure aimed at
maximising the signal--to--noise (S/N) of faint emission lines in regions away from bright
sky lines, and did not attempt optimal subtraction of the bright sky
lines themselves. Although, apparently, the data were nominally taken
in the traditional $ABBA$ sequence (Richard et al., 2003, hereafter R03)
i.e. two slit positions, for processing in pairs, in fact in all three
groups the object was placed at several positions along the slit: at 5
positions for two of the groups, and at 6 positions for the third
group. The essence of our reduction procedure was to take advantage of
this, and to achieve refined bias, dark, and sky subtraction by, for each slit
position, averaging all the frames at the other slit positions, and
subtracting. Relative to the usual procedure of processing the data in
pairs, consideration of the propagation of errors reveals that for $N$ slit
positions this procedure should reduce the noise by
a factor $\sqrt{2(N-1)/N}$, which gives 1.26 for $N=5$. In a previous analysis of 
ISAAC data (Weatherley et al. 2005, MNRAS, submitted), we verified that the improvement 
predicted by this formula is indeed achieved. 

From an analysis of dark frames we noted that the ISAAC dark varies
on timescales of a single frame by a DC offset. This needs to be
removed before flat--fielding, or the flat--field signature multiplied
by the differential DC offset will be added into the frames.
Fortunately the slit does not fully cover the array, and we used the
unexposed region to monitor this variation. The third data group
suffers from 50Hz pick--up noise. This was removed by a Fourier
procedure, identifying the relevant wave vectors in the power
spectrum, masking all other wave vectors, inverse transforming, and
subtracting.  In detail, then, the steps we followed were: 1. subtract notional
dark; 2. removal
of differential DC offset; 3. flat--field; 4. fit sky up columns;
5. create frames for residual dark+bias+sky subtraction by forming, for each slit position, 
the median of all the other processed frames at the other positions; 6. subtract; 7. remove
pick--up noise; 8. refit sky up columns. This procedure will work well
if two conditions are satisfied: the dark pattern (modulo a DC offset)
is fairly stable over each group, and the error in the flat--field
multiplied by the variation in the sky level, is small relative to the
detector noise.

At this point the ISAAC data--reduction manual recommends wavelength
calibration of the 2D frames. In one sense this is essential, in order
to register the frames, because the sky lines are curved -- meaning that
for a given spatial shift, the required spectral shift varies with
position over the frame. Wavelength calibration involves rebinning,
which has the unfortunate consequences that bad pixels become harder
to recognise (because they become smoothed out) and that covariance
between pixels is introduced. We considered it vital to keep the data
in pixels independent in order to allow an accurate estimation of the errors
(see below). Fortunately this is in fact possible because we are only
interested in a small spectral and spatial region; the region where the two
wavelength ranges overlap, at the position of the target. For each frame we established the spatial
shift by measuring the position of the bright star that was centred
in the slit (P04, fig. 1). We then established the appropriate
wavelength shift at the spatial position of the target galaxy, from sky lines at a
wavelength close to the wavelength of interest. Then, for each group,
we registered all the frames using integer pixel shifts, spectrally
and spatially. Registration will be correct in the region of interest,
but the data will be smeared at other wavelengths and spatial
locations. We then scaled the data to the same count level using the
counts detected from the bright star. We then measured the noise in a
region of low sky (akin to the region of the sought line), and finally
combined the 2D frames using inverse--variance weights. There 
are three objects of interest on the slit: the target emission
line, the bright star, and a galaxy at $z=1.68$, called $\#2582$,
(discussed in R03). The data for this galaxy presented in R03 are the
same data analysed by P04. An emission line from galaxy $\#2582$ is
visible in fig. 4 in P04. We use the
emission--line galaxy to flux calibrate our data, using the flux
quoted in R03. This ensures that we are on the same flux scale as that
of P04, and can compare the noise in our final combined frame to
theirs. At the same time we checked our wavelength calibration against theirs, by comparing the
two measurements of the wavelength of this line, finding good agreement.
The star is useful for measuring the
wavelength dependence of atmospheric absorption. To this end we repeated the registration procedures
appropriate for the two other targets, and produced combined frames
for each object.

There is another registration issue that has not been considered
above, caused by the fact that the centering of the target in the slit
may vary from frame to frame, resulting in a small wavelength shift of any
emission line\footnote{We are grateful to Roser
Pell\'{o} for bringing this to our attention.}. We quantified the
importance of this effect by measuring the variation of the wavelength
difference between the \OIII\ line from galaxy $\#2582$, and the
adjacent bright sky line. The standard deviation of this difference over
the 33 frames is $\sigma=0.66$\thinspace pixels, or 0.1\arcsec. It is easy to see
that this does not have a significant effect on the detectability of
the line. Recall that the resolution of an unresolved line is set by
the seeing, because smaller than the slit width. For 0.5\arcsec
seeing, 3.4 pixels FWHM, adding the spread of shifts in quadrature
would increase the wavelength FWHM to 3.7 pixels, an increase of only
$10\%$. Nevertheless for the sake of completeness we added these small 
increments to the previously computed wavelength shifts, in calculating the 
relevant integer pixel shifts in the registration process. The results in this paper
are with these shifts included. We also produced final frames without
these shifts included, and the relevant results are unchanged. We emphasise
that we have used exactly the same wavelength information (sky--line map and wavelength drift of the \OIII\ line)
 as P04 to register the pixels containing the claimed emission 
line. This means that the only difference in our registration procedure is our use of integer 
pixel shifts. For an image FWHM of 3.4 pixels (spectrally and spatially), we have computed that
integer pixel shifts broaden the image by only 2 per\thinspace cent.

The individual frames suffer from bad pixels and cosmic--ray hits, and
it is necessary to identify these and reject them. We experimented
with a number of rejection schemes to optimise the rejection of bad
data, and to maximise the final S/N. Our preferred scheme compared the
counts in a particular pixel to the median counts in that pixel for
all the registered frames of the group, $\sigma$-clipping bad data using the
estimated error for that pixel (see below). Nevertheless we also
produced frames by several other methods, including rejection of the
brightest and faintest pixels (IRAF {\em minmax}), percentile--clipping
(IRAF {\em pclip}), and simple medianing. Because our conclusion is
that the line is not detected, we closely inspected all these frames,
as well as smoothed versions, to search for the line and to look for
discrepancies between the different rejection methods. We found no
noticeable differences. A final frame was produced for $\lambda1$ and
for $\lambda2$, for each of the three targets, of limited spatial
extent. Finally, for each object, we registered the $\lambda1$ and
$\lambda2$ frames, again using integer spatial and spectral shifts,
scaled them to a common count level, and averaged, using
inverse--variance weights. The final frames were flux calibrated, as
explained above. We made no correction for atmospheric absorption in
this step, because inspection of the star spectrum indicates that the
degree of absorption is similar at the \lya\ wavelength, and the
wavelength of the \OIII\ line used in the calibration. The final
sub--frame for the \lya\ emission line is shown in Fig. 1.
Residuals of the 50Hz pick--up noise (wavelength in the Y direction
6.5 pixels) are detectable in this frame at a very low level, well
below the sky noise. In searching for an emission line in the final
frame, we also searched a frame in which we subtracted off the
median value of the counts in each row, which will remove any residual pick--up noise, but leave the emission line unaffected.

To assess the significance of any detection requires an accurate
estimate of the noise in each pixel. An end--to--end Poisson estimate,
accounting for read noise, is one approach. In our experience with
near--infrared arrays, this tends to somewhat underestimate the
noise. Since we are interested in a faint emission line, we need an
accurate estimate of the noise in the sky. For this purpose we
measured the standard deviation of the counts up each column of the
final frames (estimated by iteratively $\sigma-$clipping out sources),
and used this as our estimate of the noise in each sky pixel in that
column. This procedure would not be valid had we included the
wavelength calibration stage.

\begin{figure}
\begin{center}
\label{fig:frame}
\includegraphics[width=1\columnwidth]{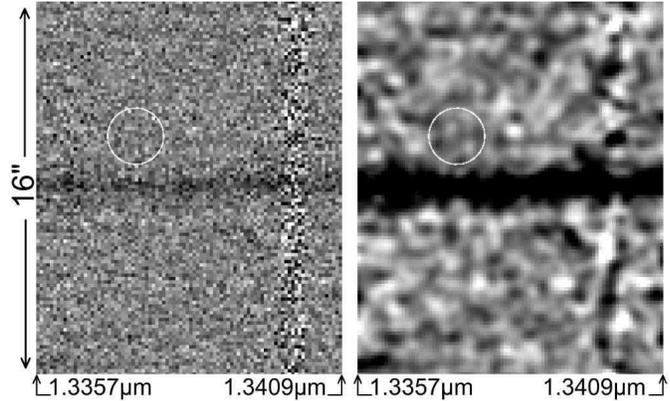}
\caption{Final combined 2D frame in a small region around the location
of the line detected by P04, marked by the circle. The wavelength
scale was computed from the wavelengths of nearby sky lines, and the
scale is assumed linear over this restricted wavelength range. No
significant line is detected at the expected location. Note that no
rebinning has occurred at any stage in the data reduction process, in
order to keep pixels independent. The dark line across the centre is emission
from the field galaxy mentioned in P04. For interest we also show the same frame
smoothed with a Gaussian kernel of $\sigma$=1.5\thinspace pixels. The data displayed here are available in the form of two FITS
frames, for the spectrum and the errors, at
http://astro.imperial.ac.uk/$\sim$sjw98/data.html}
\end{center}
\end{figure}


\section{Results}
The expected location of the \lya\ emission line is indicated in
Fig. 1 by the circle. The wavelength scale was established from the
sky lines. The emission line is not visible in this frame, neither in
a smoothed version, nor in the two sub--frames for $\lambda1$ and
$\lambda2$. The spatial FWHM in the final combined frame is
$0.5\arcsec$, or 3.4 spatial pixels. For a point source, this implies
a spectral resolution also of about 3.4 pixels. To assess the flux in the
reported line we simply integrated the flux and variance in a box of
size $5\times5$ pixels, centred on the location reported in P04, which
is $35.6\arcsec$ along the slit from galaxy $\#2582$. 
Our result at the reported location is a measured flux of
$(0.1\pm0.4)\times10^{-18}$\,erg\,s$^{-1}$\,cm$^{-2}$. By contrast
P04 reported a measured flux of
$(4.1\pm0.5)\times10^{-18}$\,erg\,s$^{-1}$\,cm$^{-2}$. However, we note that the
distance between the two objects as computed from the quoted
coordinates (P04, R03) is $35.9\arcsec$, a difference
of 2 pixels. Furthermore the optimal length of the box in the spectral
direction depends on the intrinsic line width, whether resolved or
not. For these reasons we repeated our measurements for smaller ($3\times 3$ pixels) and
larger ($7\times 7$ pixels) boxes, and searched the entire wavelength range 
visible in Fig. 1, 1.3357$\mu$m -- 1.3409$\mu$m,
shifting the centre from the nominal
location by up to 0.5 arcsec, up and down the slit, to be certain we did not miss the line. 
Over this entire region we found no emission line above $3\sigma$ significance.

Unfortunately P04 do not quote the aperture used for their measurement
of the flux, so we cannot make a direct comparison of the errors. If
the aperture sizes are similar, then a direct comparison is valid, in
which case it is interesting to note that our quoted error is slightly
smaller than theirs, by the factor predicted in the previous section.

Before considering the origin of the discrepancy in the measured
fluxes, it is interesting to compare our results for the galaxy
$\#2582$ with those of R03, measured from the same data. The galaxy
lies at a redshift $z=1.68$, confirmed by the detection of three
lines, \OIII 4959, 5007, and H$\beta$. The H$\beta$ line is the
weakest, for which R03 quote a measured flux of
$(6.6\pm1)\times10^{-18}$\,erg\,s$^{-1}$\,cm$^{-2}$.  We also detect
this line. Since this line is evidently real, and of comparable quoted
S/N to that of the \lya\ line, it must be considered surprising that
we do not also detect the \lya\ line if it is also real.

To find the cause of the discrepancy between our results
for the \lya\ line and those reported by P04, we re--reduced the data following the principles of P04, i.e.
subtracting frames in pairs, then wavelength calibrating the frames, rebinning onto a linear wavelength scale. In this process
we made a careful check for bad data. We identified three variable hot pixels\footnote{These have coordinates
(28,761), (28, 836), (919, 790) in the raw frames.} which result  in spurious positive flux in 
four of the sky--subtracted frames in the region
of the emission line. We confirmed that these are very easily identified when the frames are
registered to the nearest pixel, but are harder to spot when the data are rebinned in the wavelength
calibration step. The summed spurious positive flux, when averaged into the entire data set, corresponds approximately to the flux
measured by P04; therefore these variable hot pixels plausibly account for the difference between our results
and those of P04. Nevertheless, we hesitate to conclude that we have found the cause of the discrepancy, since
in the absence of further details of their reduction process (linear wavelength solution, pixel rejection scheme,
frame weights) we are unable to reproduce their results exactly. At the same time this analysis highlights the dangers
of rebinning near--infrared array data. Use of a bad--pixel mask is another useful approach to this problem.

Subsequent to the submission of this paper we learnt of the results of Bremer et al. (2004) who re-observed
the field, obtaining an image in the H band reaching approximately 1 mag. deeper than the observations of P04. 
The reported galaxy is not detected in this image. Taken with our non-detection of the reported emission
line, a consistent interpretation is that the galaxy does not exist, and that the original 
reported H ($4 \sigma$) and K ($3 \sigma$) detections 
are chance superpositions of statistical fluctuations in the background sky.

\begin{acknowledgements}
We are grateful to the anonymous referee for raising issues that have helped clarify 
this manuscript. 
\end{acknowledgements}

\end{document}